\newlength{\absize}
\documentclass[12pt]{article}

\usepackage{psfig}
\usepackage{epsfig}
\usepackage{dsfont}

\usepackage{graphicx}
\usepackage{color}

\usepackage{amssymb}
\usepackage{hyperref}

\renewcommand{\baselinestretch}{1.5}

\begin{document}
\thispagestyle{empty} \pagestyle{empty}
\renewcommand{\thefootnote}{\fnsymbol{footnote}}
\newcommand{\starttext}{\newpage\normalsize
\pagestyle{plain} \setlength{\baselineskip}{3ex}\par \setcounter{footnote}{0}
\renewcommand{\thefootnote}{\arabic{footnote}}
}

\newcommand{\preprint}[1]{\begin{flushright}
\setlength{\baselineskip}{3ex}#1\end{flushright}}
\renewcommand{\title}[1]{\begin{center}\LARGE
#1\end{center}\par}
\renewcommand{\author}[1]{\vspace{2ex}{\Large\begin{center}
\setlength{\baselineskip}{3ex}#1\par\end{center}}}
\renewcommand{\thanks}[1]{\footnote{#1}}
\renewcommand{\abstract}[1]{\vspace{2ex}\normalsize\begin{center}
\centerline{\bf Abstract}\par\vspace{2ex}\parbox{\absize}{#1
\setlength{\baselineskip}{2.5ex}\par}
\end{center}}

\newlength{\eqnparsize}
\setlength{\eqnparsize}{.95\textwidth}

\setlength{\jot}{1.5ex}
\newcommand{\figsize}{\small}
\renewcommand{\bar}{\overline}
\font\fiverm=cmr5
\input prepictex
\input pictex
\input postpictex
\input{psfig.sty}
\newdimen\tdim
\tdim=\unitlength

\setcounter{bottomnumber}{2} \setcounter{topnumber}{3} \setcounter{totalnumber}{4}
\renewcommand{\bottomfraction}{1}
\renewcommand{\topfraction}{1}
\renewcommand{\textfraction}{0}

\def\draft{
\renewcommand{\label}[1]{{\quad[\sf ##1]}}
\renewcommand{\ref}[1]{{[\sf ##1]}}
\renewenvironment{thebibliography}{\section*{References}}{}
\renewcommand{\cite}[1]{{\sf[##1]}}
\renewcommand{\bibitem}[1]{\par\noindent{\sf[##1]}}
}

\def\theequation{\thesection.\arabic{equation}}
\preprint{hep-th/0411256}
\title{Quantum Geometries of $A_{2}$}
\author{Girma Hailu\thanks{hailu@physics.harvard.edu}
\\
\small\sl Jefferson Laboratory of Physics\\
\small\sl Harvard University \\
\small\sl Cambridge, MA 02138 }


\abstract{We solve $\mathcal{N}=1$ supersymmetric $A_{2}$ type
$U(N)\times U(N)$ matrix models obtained by deforming
$\mathcal{N}=2$ with symmetric tree level superpotentials of any
degree exactly in the planar limit. These theories can be
geometrically engineered from string theories by wrapping D-branes
over Calabi-Yau threefolds and we construct the corresponding exact
quantum geometries.}

\starttext


\setcounter{equation}{0}
\section{Introduction\label{intr}}

A class of supersymmetric gauge theories with tree level
superpotentials can be geometrically engineered from type $IIA$ and
type $IIB$ string theories by wrapping D-branes over various cycles
of Calabi-Yau threefolds. See \cite{GV-1,V-1,KKLM-1,CIV-1,CFIKV} for
instance. The quantum Calabi-Yau geometries can be studied using the
related geometrically engineered gauge theories. Constructing the
exact quantum geometries associated to product gauge group theories
with general tree level superpotentials is a highly nontrivial
problem. More recently, important connections between matrix models
and supersymmetric gauge theories have been found by Dijkgraaf and
Vafa. \cite{DV-1a,DV-1,DV-1b} It was also found by expansion
\cite{DV-1} that the quantum Calabi-Yau geometry that engineers
$A_{2}$ could be expressed in terms of two polynomials. In this
note, we will give analytic proof and find these polynomials by
solving the matrix models for $\mathcal{N}=1$ supersymmetric $A_{2}$
type $U(N)\times U(N)$ gauge theories with symmetric tree level
superpotentials of any degree in the planar limit and thus construct
the corresponding exact quantum Calabi-Yau geometries.

First let us start with a general $\mathcal{N}=2$ supersymmetric
$\prod_{i}U(N_i)$ gauge theory with link chiral superfields $Q_{ij}$
and $Q_{ji}=Q_{ij}^{\dagger}$ transforming as $(\Box,\,\bar{\Box})$
and $(\bar{\Box},\,\Box)$ respectively under $U(N_{i})\times
U(N_{j})$ and the corresponding matrix model. Consider the tree
level superpotential
\begin{equation} W_{\mathrm{tree}}(\Phi,Q)=\sum_{i,j}s_{ij}\mathrm{Tr }\,
Q_{ij}\Phi_{j}Q_{ji}+\sum_{i}\mathrm{Tr }\,
W_{i}(\Phi_{i}),\label{eq:n2-1}\end{equation} where $\Phi_{i}$ is
the scalar chiral superfield associated with $U(N_i)$ and the
indices are ordered such that $s_{ij}=-s_{ji}=1$ with $j>i$ when the
$i^{th}$ and $j^{th}$ gauge groups are linked and $s_{ij}=0$
otherwise. The first term in (\ref{eq:n2-1}) comes from the
superpotential of $\mathcal{N}=2$ with bifundamental
hypermultiplets. The second term is a polynomial in each $\Phi_{i}$
and it will contain quadratic mass terms which break $\mathcal{N}=2$
down to $N=1$. This theory can be geometrically engineered from type
IIB string theory with D3, D5 and D7 branes wrapped over various
cycles of Calabi-Yau threefolds and also from type IIA string theory
with D6 branes wrapped over Calabi-Yau threefolds.
\cite{GV-1,V-1,KKLM-1,CIV-1,CFIKV} The partition function is defined
as
\begin{equation}
Z=\frac{1}{Z_{0}}\int
\prod_{i}d\Phi_{i}\prod_{i<j}dQ_{ij}dQ_{ji}\,e^{-\frac{1}{g_{s}}
W_{\mathrm{tree}}(\Phi,Q)}.\label{zmoose}
\end{equation}
and normalized such that in terms of the eigenvalues
$\lambda_{i,1},\cdots,\lambda_{i,I}, \cdots\lambda_{iN_{i}}$ of
$\Phi_{i}$ it becomes
\begin{equation}
Z=\int\prod_{i,I}d\lambda_{i,I}\mathrm{exp}(-S_{\mathrm{eff}}),
\label{eq:zpq-2}
\end{equation}
with the effective action \cite{K-1}
\begin{equation}
S_{\mathrm{eff}}=\frac{1}{g_{s}}\sum_{i,I}W_{i}(\lambda_{i,I})
-2\sum_{i,I<J}\mathrm{log}|\lambda_{i,I}-\lambda_{i,J}|
+\sum_{i<j,I,J}|s_{ij}|\mathrm{log}|\lambda_{i,I}
-\lambda_{j,J}|.\label{eq:zpq-3}
\end{equation}
Note that the small letter index $i$ denotes the $i^{\mathrm{th}}$
gauge group and the upper letter index $I$ denotes eigenvalues. The
equations of motion are obtained by minimizing (\ref{eq:zpq-3}) with
$\lambda_{i,I}$,
\begin{equation}
W'_{i}(\lambda_{i,I})-2g_{s}\sum_{J\ne
I}\frac{1}{\lambda_{i,I}-\lambda_{i,J}}+g_{s}\sum_{j,J}|s_{ij}|
\frac{1}{\lambda_{i,I}-\lambda_{j,J}}=0.\label{eq:zpq-4}
\end{equation}
Let us introduce the resolvents,
\begin{equation}
w_{i}(x)=\frac{1}{N_{i}}\sum_{I=1}^{N_{i}}\frac{1} {\lambda_{i,I}-x},\label{eq:zpq-5}
\end{equation}
where $x$ is complex. Note that $w_{i}(x)$ obey the asymptotic
larger $x$ behavior
\begin{equation}
w_i(x)\to -\frac{1}{x}.\label{wasymp}
\end{equation}
The eignevalues are distributed on the real
axis of $x$. We will consider the case in which the gauge symmetry
in the low energy theory is unbroken in this note. This corresponds
to the case in which each set of eignevalues $\lambda_{i,I}$ is
separately distributed on a single interval $[a_{i},\,b_{i}]$. The
equations of motion (\ref{eq:zpq-4}) expressed in terms of the
resolvents give
\begin{equation}
S_{i}(w_{i}(x+i0)+w_{i}(x-i0))-\sum_{j}|s_{ij}|S_{j}w_{j}(x)+W'_{i}(x)=0
\label{resqe}
\end{equation}
for $x\in [a_{i},\,b_{i}]$ where
\begin{equation}
S_{i}\equiv g_{s}N_{i}.\label{sing}
\end{equation}
Following Dijkgraaf and Vafa \cite{DV-1a,DV-1,DV-1b}, $S_i$ will be
identified with the glueball superfields  defined in terms of the
gauge chiral superfields $W_{i\alpha}$ associated to the confining
$SU(N_{i})$ subgroup of  $U(N_{i})$ as
\begin{equation}
S_{i}=-\frac{1}{32\pi^2}\mathrm{Tr\,}W_{i}^{\alpha}W_{i\alpha}.\label{eq:rev4-2}
\end{equation}
In the large $N$ limit,  the eigenvalues are continuously
distributed and each resolvent $w_i(x)$ can be written as a sum a
regular function $w_{ir}(x)$ which is a particular solution of
(\ref{resqe}) and another function $w_{is}(x)$ which contains the
singular part of $w_{i}(x)$,
\begin{equation}
w_{i}(x)=w_{ir}(x)+w_{is}(x). \label{resrs}
\end{equation}
We can think of this as a substitution for $w_{i}(x)$ in terms of
$w_{ir}(x)+w_{is}(x)$ where $w_{ir}(x)$ satisfies the regular
equation (\ref{zpq4s}) below and we will then solve for $w_{is}(x)$
such that the asymptotic behavior (\ref{wasymp}) is satisfied. We
will find that $w_{is}(x)$ is the singular part of the resolvent.

Putting (\ref{resrs}) in (\ref{resqe}) and setting
\begin{equation}
2S_{i}w_{ir}(x)-\sum_{j}|s_{ij}|S_{j}w_{jr}(x)+W'_{i}(x)=0,\label{zpq4s}
\end{equation}
we obtain
\begin{equation}
S_{i}(w_{is}(x+i0)+w_{is}(x-i0))-\sum_{j}|s_{ij}|S_{j}w_{js}(x)=0,\label{resqes}
\end{equation}
for $x$ in the branch cut $[a_i,\,b_i]$ of $w_{is}(x)$. In the large
$N$ limit, the eigenvalues become continuous and we introduce the
eigenvalue densities
\begin{equation}
\rho_{i}(\lambda)=\frac{1}{N_{i}}\sum_{I}
\delta(\lambda-\lambda_{i,I}),\label{eq:mrho-1}
\end{equation}
normalized such that $\int \rho_{i}(\lambda)d\lambda=1$, and
(\ref{eq:zpq-5}) becomes
\begin{equation} w_{i}(x)=\int\frac{\rho_{i}
(\lambda)d\lambda}{\lambda-x}.\label{wixr}
\end{equation}
Once $w_i(x)$ are found, (\ref{wixr}) can be inverted to determine
$\rho_{i}(\lambda)$ and
\begin{equation}
\rho_{i}(\lambda)=\frac{1}{2\pi i}(w_i(\lambda+i0)-w_i(\lambda-i0)).
\label{wiri}
\end{equation}

The multi-matrix planar free energy can be conveniently written as
\begin{equation} \mathcal{F}_{0}  =
\frac{1}{2}\sum_{i}S_{i}\int d\lambda\rho_{i}
(\lambda)W_{i}(\lambda)-\frac{1}{2}\sum_{i,j}C_{ij}S_{i}S_{j}\int
d\lambda\rho_{i} (\lambda)\mathrm{log}|\lambda|\,,\label{f0rs}
 \end{equation}
where $C_{ij}=2\delta_{ij}-|s_{ij}|$ is the Cartan matrix. We will
not do free energy calculations in this note. The reason we have
added this last paragraph is because we find the free energy given
by (\ref{f0rs}) in terms of single integrals simpler and useful for
doing calculations and we have not seen it in the literature on
multi-matrix models. The derivation is given in Appendix
\ref{appfii}.

\setcounter{equation}{0}
\section{Quantum geometries of $A_{2}$}\label{sqga2}

In this section we will explicitly construct the quantum Calabi-Yau
geometries associated to $\mathcal{N}=1$ supersymmetric $A_{2}$ type
$U(N)\times U(N)$ gauge theories obtained by deforming
$\mathcal{N}=2$ with symmetric tree level superpotentials of any
degree and the gauge symmetry unbroken in the low energy theory. In
the low energy theory, the $U(1)$ subgroup of each $U(N)$ decouples
and the $SU(N)$ subgroup confines. The most general asymptotically
free product gauge theories of the type discussed in Section
\ref{intr} for the confining $\Pi_{i} SU(N_{i})$ subgroup with
$\mathcal{N}=2$ supersymmetry and link chiral superfields in the
bifundamental representation are constrained to be only of $A-D-E$
type Dynkin diagrams. See \cite{KMV} for instance. The reason is
that the condition of asymptotic freedom for the $i^\mathrm{th}$
gauge group can be written as $(2\delta_{ij}-\sum_{j\neq
i}|s_{ij}|)N_{j}>0$ and this results in the constraint that all
eigenvalues of the connectivity matrix $|s_{ij}|$ need to be less
that $2$ in order for the theory to be asymptotically free. Thus
$(2\delta_{ij}-|s_{ij}|)$ is the Cartan matrix of $A-D-E$ type
Dynkin diagrams and the most general asymptotically free such
$\mathcal{N}=2$ product gauge theories with link chiral superfields
in the bifundamental representation are of $A-D-E$ type. When the
eignevalues of the connectivity matrix also contain $2$, the beta
function vanishes and theory is conformal and the diagram is that of
affine $\hat{A}-\hat{D}-\hat{E}$ type.

Our interest is $\mathcal{N}=2$ supersymmetric $A_2$ type
$U(N)\times U(N)$ gauge theory deformed to $\mathcal{N}=1$ by
symmetric tree level superpotentials with the gauge symmetry
preserved in the low energy theory. This corresponds to two separate
cuts for the resolvents associated to each gauge group in the matrix
model. It follows from (\ref{wiri}) that each branch cut is a square
root branch cut. The regular parts of the resolvents $w_{1r}(x)$ and
$w_{2r}(x)$ are solutions of the following equations which follow
from (\ref{zpq4s}) with $i,j$ running over $1,2$,
\begin{equation}
2S_{1}w_{1r}(x)-S_{2}w_{2r}(x)+W'_{1}(x)=0\,, \quad
2S_{2}w_{2r}(x)-S_{1}w_{1r}(x)+W'_{2}(x)=0. \label{a2-19}
\end{equation}
The solutions are
\begin{equation}
w_{1r}(x)=-\frac{1}{3S_{1}}\Bigl(2W'_{1}(x)+W'_{2}(x)\Bigl)\,,\quad
w_{2r}(x)=-\frac{1}{3S_{2}}\Bigl(2W'_{2}(x)+W'_{1}(x)\Bigl)\label{a2-19a}
\end{equation}
Now the tree level superpotential (\ref{eq:n2-1}) becomes
\begin{equation} W_{\mathrm{tree}}(\Phi,Q)=\mathrm{Tr }\,
Q_{12}\Phi_{2}Q_{21}-\mathrm{Tr }\, Q_{21}\Phi_{1}Q_{12}+\mathrm{Tr
}\, W_{1}(\Phi_{1})+\mathrm{Tr }\, W_{2}(\Phi_{2}).\label{eq:n2-a2}
\end{equation}
The classical equations of motion are
\begin{eqnarray}
Q_{12}\Phi_{2}-\Phi_{1}Q_{12}=0,\quad
Q_{21}\Phi_{1}-\Phi_{2}Q_{21}=0,\nonumber \\
-Q_{12}Q_{21}+\frac{\partial{W_{1}(\Phi_{1})}}{\partial{\Phi_{1}}}=0,\quad
Q_{21}Q_{12}+\frac{\partial{W_{2}(\Phi_{2})}}{\partial{\Phi_{2}}}=0.\label{a2class-1}
\end{eqnarray}
Combining these equations, we can write
\begin{eqnarray}
(X-S_{1}w_{1r})(X+S_{1}w_{1r}-S_{2} w_{2r})=0,\quad \quad (X+S_{2}w_{2r})=0,\nonumber\\
(Y-S_{2}w_{2r})(Y-S_{1}w_{1r}+S_{2}w_{2r})=0,\quad \quad
(Y+S_{1}w_{1r})=0, \label{a2class-2}
\end{eqnarray}
where $X=-Q_{21}Q_{12}-S_{1}w_{1r}+S_{2} w_{2r}$ and
$Y=Q_{12}Q_{21}+S_{1}w_{1r}-S_{2} w_{2r}$. The singular classical
spectral curve can be written in terms of a complex variable $y$ as
\begin{equation}
(y+S_{1}w_{1r}(x))(y-S_{2}w_{2r}(x))(y-S_{1}w_{1r}(x)+S_{2}
w_{2r}(x))=0. \label{class-sc1}
\end{equation}
The corresponding classical Calabi-Yau geometry is the singular
threefold,
\begin{equation}
uv+(y+S_{1}w_{1r}(x))(y-S_{2}w_{2r}(x))(y-S_{1}w_{1r}(x)+S_{2}
w_{2r}(x))=0, \label{class-cy1}
\end{equation}
which describes the $A_2$ fibration over the $x$ plane, where $u$,
$v$ and $y$ are complex coordinates. At the quantum level, the
classical singularities are resolved and the spectral curve that
describes the quantum resolution of the geometry is that of the
resolved threefold and it should be given by (\ref{class-sc1}) with
the classical values of the resolvents replaced by the singular
parts of the quantum resolvents,
\begin{eqnarray}
&&(y+S_{1}w_{1s}(x)))(y-S_{2}w_{2s}(x))(y-S_{1}w_{1s}(x)+S_{2}
w_{2s}(x))\nonumber\\
&&=(y-S_{1}w_{1r}(x)+S_{1}w_{1}(x)))(y+S_{2}w_{2r}(x)-S_{2}w_{2}(x))\nonumber\\
&& (y+S_{1}w_{1r}(x)-S_{2} w_{2r}(x)-S_{1}w_{1}(x)+S_{2}
w_{2}(x))=0. \label{quantum-sc1}
\end{eqnarray}
Putting the decomposition given by (\ref{resrs}) in
(\ref{quantum-sc1}) and using the classical solution given by
(\ref{a2-19a}) gives
\begin{equation}
(y+S_{1}w_{1r}(x))(y-S_{2}w_{2r}(x))(y-S_{1}w_{1r}(x)+S_{2}
w_{2r}(x))-f(x)\,y-g(x)=0, \label{quantum-sc1b}
\end{equation}
where
\begin{eqnarray}
f(x)&=&S_{1}^{2}w_{1s}(x)^{2}+S_{2}^{2}w_{2s}(x)^{2}
-S_{1}S_{2}w_{1s}(x)w_{2s}(x)\nonumber\\
&&-\frac{1}{3}(W_{1}'(x)^2+
W_{2}'(x)^2+W_{1}'(x)W_{2}'(x))\label{quantum-f}
\end{eqnarray}
and
\begin{eqnarray}
g(x)&=&S_{1}^2 S_{2} w_{1s}(x)^{2}w_{2s}(x)-S_{1} S_{2}^{2}
w_{1s}(x)
w_{2s}(x)^{2}\nonumber\\
&&-\frac{1}{27}(2W_{1}'(x)^{3}-2W_{2}'(x)^{3}
+3W_{1}'(x)^{2}W_{2}'(x)-3W_{1}'(x)W_{2}'(x)^{2}).\label{quantum-g}
\end{eqnarray}

The relation between Calabi-Yau geometries and matrix models was
first found in \cite{DV-1a,DV-1,DV-1b} and the general classical and
quantum curves as in the form (\ref{class-sc1}) and
(\ref{quantum-sc1b}) in terms of two general polynomial functions
$f(x)$ and $g(x)$ was given in \cite{DV-1}. These polynomials
describe the resolution of the singularities in the quantum theory
and constructing them for general tree level superpotential is a
highly nontrivial problem. Here we will construct the exact
polynomials $f(x)$ and $g(x)$ that describe the quantum resolved
geometry for $U(N)\times U(N)$ with symmetric tree level
superpotentials $W_{1}(x)$ and $W_{2}(x)=W_{1}(-x)$ of any degree
with the gauge symmetry preserved in the low energy theory.

The singular parts of the resolvents  satisfy (\ref{resqes}) which
for the case of $U(N_1)\times U(N_2)$ becomes
\begin{equation}
S_{1}(w_{1s}(x+i0)+w_{1s}(x-i0))-S_{2}w_{2s}(x)=0\quad \mathrm{for}\, x\in [a_1,\,
b_1]\label{a2-1}
\end{equation}
and
\begin{equation}
S_{2}(w_{2s}(x+i0)+w_{2s}(x-i0))-S_{1}w_{1s}(x)=0\quad \mathrm{for}\, x\in [a_2,\,
b_2].\label{a2-2}
\end{equation}
The resolvents satisfy two independent equations, one quadratic and
the other cubic,
\begin{eqnarray}
S_{1}^{2}w_{1}(x)^{2}+S_{2}^{2}w_{2}(x)^{2}
-S_{1}S_{2}w_{1}(x)w_{2}(x)+S_{1}W_{1}'(x)w_{1}(x)\nonumber \\
 +S_{2}W_{2}'(x)w_{2}(x)
+f_{1}(x) +f_{2}(x)=0\label{w12quad}
\end{eqnarray}
and
\begin{eqnarray}
S_{1}^2 S_{2} w_{1}(x)^{2}w_{2}(x)-S_{1} S_{2}^{2} w_{1}(x)
w_{2}(x)^{2}+S_{1}^{2} w_{1}(x)^2W_{1}'(x)+S_{1}w_{1}(x)W_{1}'(x)^2
+f_{1}(x)W_{1}'(x)\nonumber\\
-g_{1}(x)-S_{2}^{2}w_{2}(x)^2
W_{2}'(x)-S_{2}w_{2}(x)W_{2}'(x)^2-f_{2}(x)W_{2}'(x)+g_{2}(x)=0,
\label{w12cub}
\end{eqnarray}
where
\begin{equation}
f_{i}(x)\equiv\frac{S_{i}}{N_{i}}\sum_{I}\frac{W_{i}'(x)
-W_{i}'(\lambda_{iI})}{x-\lambda_{i,I}}\label{aq11}
\end{equation}
and
\begin{equation}
g_{1}(x)\equiv\frac{S_{1}S_{2}}{N_{1}N_{2}}\sum_{I,J}\frac{W_{1}'(x)
-W_{1}'(\lambda_{1I})}{(\lambda_{1,I}-\lambda_{2,J})(x-\lambda_{1,I})},\quad
g_{2}(x)\equiv\frac{S_{1}S_{2}}{N_{1}N_{2}}\sum_{I,J}\frac{W_{2}'(x)
-W_{2}'(\lambda_{2I})}{(\lambda_{2,I}-\lambda_{1,J})(x-\lambda_{2,I})}.
\label{aq11b}
\end{equation}
are polynomials. The quadratic equation for the $O(n)$ matrix model
was obtained in \cite{EKZ} and the quadratic and cubic equations
(\ref{w12quad}) and (\ref{w12cub}) for the $A_2$ model were obtained
in \cite{L-1,KLLRNSW,NSY-1,CT-1}. We have also given a derivation in
Appendix \ref{appqce}. The most general independent equations that
$w_{1s}(x)$ and $w_{2s}(x)$ satisfy are at most cubic in either
$w_{1s}(x)$ and $w_{2s}(x)$ or their combinations and there are
three Riemann sheets with one cut for $w_{1s}(x)$ joining the first
and the second sheets and a second cut for $w_{2s}(x)$ joining the
second and third sheets.

In the large $N$ limit, $w_{i}(x)$ can be expressed as in
(\ref{resrs}) which with (\ref{a2-19a}) in (\ref{w12quad}) and
(\ref{w12cub}) gives
\begin{equation}
S_{1}^{2}w_{1s}(x)^{2}
+S_{2}^{2}w_{2s}(x)^{2}-S_{1}S_{2}w_{1s}(x)w_{2s}(x)=3p(x)\label{wb12a}
\end{equation}
and
\begin{equation}
S_{1}^{2}S_{2}w_{1s}(x)^{2}w_{2s}(x)
-S_{1}S_{2}^{2}w_{1s}(x)w_{2s}(x)^{2} =2q(x),\label{wb12b}
\end{equation}
where
\begin{equation}
p(x)=\frac{1}{9}\Bigl(W_{1}'(x)^2+W_{2}'^2+W_{1}(x)'W_{2}'(x)
-3f_{1}(x)-3f_{2}(x)\Bigr)\label{ppol}
\end{equation}
and
\begin{eqnarray}
q(x)&=&\frac{4}{27}((W'_{1}(x)^{3}-W'_{2}(x)^{3})
+\frac{1}{18}(W'_{1}(x)^{2}W'_{2}(x)-W'_{1}(x)W'_{2}(x)^{2})\nonumber\\
&&-\frac{1}{2}(W'_{1}(x)f_{1}(x)-W'_{2}(x)f_{2}(x))
-((W'_{1}(x)-W'_{2}(x))p(x) +\frac{1}{2}(g_{1}(x)-g_{2}(x)).
\label{qpol}
\end{eqnarray}
Using (\ref{wb12a})-(\ref{qpol}) in (\ref{quantum-f}) and
(\ref{quantum-g}) we have
\begin{equation}
f(x)=3p(x)-\frac{1}{3}(W_{1}'(x)^2+
W_{2}'(x)^2+W_{1}'(x)W_{2}'(x))\label{quantum-f1}
\end{equation}
and
\begin{equation}
g(x)=2q(x)-\frac{1}{27}(2W_{1}'(x)^{3}-2W_{2}'(x)^{3}
+3W_{1}'(x)^{2}W_{2}'(x)-3W_{1}'(x)W_{2}'(x)^{2}).\label{quantum-g1}
\end{equation}
Note that $p(x)$ is a polynomial of degree $2n_{1}$ or $2n_{2}$ and
$q(x)$ is a polynomial of degree $3n_{1}$ or $3n_{2}$ depending on
whether $n_{1}>n_{2}$ or not. Combining (\ref{wb12a}) and
(\ref{wb12b}) gives
\begin{equation}
S_{1}^{3}w_{1s}(x)^{3}-3p(x)S_{1}w_{1s}(x)=
-S_{2}^{3}w_{2s}(x)^{3}+3p(x)S_{2}w_{2s}(x)=2q(x),\label{wb12}
\end{equation}
Our interest is the case in which the two gauge groups are the same
and the potentials $W_{1}(x)$ and $W_{2}(x)$ have the same degree
$n+1$ with $n_{1}=n_{2}=n$ and are symmetric about the origin such
that $W_{2}(x)=W_{1}(-x)$. We will set $N_1=N_2=N$ from now on and
the gauge symmetry is $U(N)\times U(N)$ and it will be preserved in
the low energy theory. Thus we also have $S_1=S_2=g_{s}N\equiv S$.
This corresponds to two separate cuts in $w_{1}(x)$ and $w_{2}(x)$
associated to each gauge group in the matrix model. We will set up
the potentials $W_1(x)$ and $W_2(x)$ such that the branch cuts of
$w_{1}(x)$ and $w_{2}(x)$ will be symmetrically on the positive and
the negative real axis of $x$ respectively and we have
$\lambda_{1,I}>0$ and $\lambda_{2,J}<0$. The equations that
$Sw_{1s}(x)$ and $-Sw_{2s}(x)$ satisfy are similar to that of the
$O(n)$ matrix model investigated in \cite{EKZ} and we will use
techniques developed in \cite{EKZ} to solve the $U(N)\times U(N)$
matrix model. We will impose appropriate boundary conditions that
produce the desired properties described above. First we start with
one of the solutions to the cubic equation (\ref{wb12}) for
$w_{1s}(x)$,
\begin{equation}
w_{1s}(x)=\frac{1}{S}\Bigl(e^{-2\pi i/3} w_{s+}(x) +e^{2\pi
i/3}w_{s-}(x)\Bigr),\label{w1bs}
\end{equation}
where
\begin{equation}
w_{s\pm}(x)=\Bigl(q(x)\mp
\sqrt{q(x)^2-p(x)^3}\Bigr)^{1/3}.\label{wpmg}
\end{equation}
It follows from (\ref{wpmg}) that
\begin{equation}
p(x)=w_{s+}(x)w_{s-}(x), \label{pwpwm}
\end{equation}
\begin{equation}
q(x)=\frac{1}{2}\Bigl(w_{s+}(x)^{3}+w_{s-}(x)^{3}\Bigl),
\label{qwqwm}
\end{equation}
and
\begin{equation}
\sqrt{q(x)^2-p(x)^3}=\frac{1}{2}\Bigl(w_{s+}(x)^{3}
-w_{s-}(x)^{3}\Bigl).\label{sqdw}
\end{equation}
The second resolvent $w_{2s}(x)$ also follows as one of the three
solutions to the cubic equation in (\ref{wb12}) with appropriate
boundary conditions to be imposed,
\begin{equation}
w_{2s}(x)=\frac{1}{S}\Bigl(e^{-\pi i/3} w_{s+}(x) +e^{\pi
i/3}w_{s-}(x)\Bigr).\label{w2bs}
\end{equation}
The third solution to the cubic equation is a linear combination of
the two resolvents.

The square root branch cuts in $w_{1s}(x)$ and $w_{2s}(x)$ come from
$\sqrt{q(x)^2-p(x)^3}$. In order to fulfill the constraint that the
two branch cuts be symmetric, we need to have $a_2=-b_1$ and
$b_2=-a_1$ so that $w_{1s}(x)$ will have its branch cut on
$x\in[a\,,b]$ and $w_{2s}(x)$ on $x\in[-b\,,-a]$ with $b>a>0$. This
will be achieved by imposing the following symmetries which are
extensions of the symmetries imposed in \cite{EKZ} for the $O(n)$
matrix model,
\begin{equation}
w_{s\pm}(x-i0)=e^{\pm 2\pi i/3}w_{s\mp}(x+i0)\quad \mathrm{for} \,
x\in [a,\,b], \label{w1barbc}
\end{equation}
\begin{equation}
w_{s\pm}(x-i0)=e^{\pm 4\pi i/3}w_{s\mp}(x+i0)\quad \mathrm{for} \,
x\in [-b,\,-a] \label{w2barbc}
\end{equation}
\begin{equation}
w_{s+}(x)=w_{s-}(-x). \label{wpmconst}
\end{equation}
It then follows from (\ref{w1bs}), (\ref{wpmg}), (\ref{w2bs}) and
(\ref{wpmconst}) that $w_{2s}(x)=-w_{1s}(-x)$. Note also that
combining (\ref{wpmconst}) with (\ref{sqdw}) implies that
$q(0)^2=p(0)^3$. The main reason for the choice of the symmetries
(\ref{w1barbc}) - (\ref{wpmconst}) on $w_{1s}(x)$ and $w_{2s}(x)$
given in (\ref{w1bs}) and (\ref{w2bs}) is that we have for
$x\in[a\,,b]$,
\begin{equation}
w_{1s}(x-i0)=\frac{1}{S}(w_{s+}(x+i0)+w_{s-}(x+i0)),\quad\quad
w_{2s}(x-i0)=w_{2s}(x+i0), \label{w1sabb}
\end{equation}
and for $x\in[-b\,,-a]$,
\begin{equation}
w_{2s}(x-i0)=-\frac{1}{S}(w_{s+}(x+i0)+w_{s-}(x+i0)),\quad\quad
w_{1s}(x-i0)=w_{1s}(x+i0). \label{w2smabb}
\end{equation}
Thus $w_{1s}(x)$ has a branch cut across $x\in[a\,,b]$ and no
discontinuity across $x\in[-b\,,-a]$. On the other hand, $w_{2s}(x)$
has a branch cut across $x\in[-b\,,-a]$ and no branch cut across
$x\in[a\,,b]$. This is exactly what we wanted.

It follows from (\ref{w1bs}), (\ref{w2bs}) and (\ref{wpmconst}) that
\begin{equation}
w_{s\pm}(x)=-\frac{iS}{\sqrt{3}}\Bigl(e^{-2\pi i/3} w_{1s}(\pm x) -
e^{2\pi i/3}w_{1s}(\mp x)\Bigr)=\frac{iS}{\sqrt{3}}\Bigl(e^{-\pi
i/3} w_{2s}(\pm x) - e^{\pi i/3}w_{2s}(\mp x)\Bigr).\label{w1ps1s}
\end{equation}
The asymptotic behavior of $w_{i}(x)$ given by (\ref{wasymp})
combined with (\ref{resrs}) and (\ref{w1ps1s}) gives the asymptotic
large $x$ behaviors
\begin{equation}
w_{\pm}(x)\to\mp\frac{iS}{\sqrt{3}x},\label{w1ps1sasy}
\end{equation}
where
\begin{equation}
w_{\pm}(x)=w_{r\pm}(x)+w_{s\pm}(x)\label{wpmrsa}
\end{equation}
and $w_{r\pm}(x)$ are given by the same expressions given in
(\ref{w1ps1s}) for $w_{s\pm}(x)$ with $w_{is}(\pm x)$ replaced by
$w_{ir}(\pm x)$. Noting that $Sw_{r\pm}(x)$ are independent of $S$,
let us define
\begin{equation}
\Omega_{\pm}(x)\equiv\frac{\partial
\,(Sw_{\pm})}{\partial{S}}=\frac{\partial
\,(Sw_{s\pm})}{\partial{S}}.\label{omegadef}
\end{equation}
It is then convenient to decompose $w_{s\pm}(x)$ as
\begin{equation}
w_{s\pm}(x)=\Omega_{\pm}(x)h_{\pm}(x),\label{w1spmdec}
\end{equation}
with $\Omega_{\pm}(x)$ obeying the same boundary conditions
(\ref{w1barbc}) - (\ref{wpmconst}) as $w_{s\pm}(x)$ and having the
same large $x$ asymptotic behaviors given in (\ref{w1ps1sasy}) for
$w_{\pm}(x)$. Note that because $\Omega_{\pm}(x)$ are obtained by
taking first derivative of $Sw_{s\pm}(x)$ which have square root
branch cuts, $\Omega_{\pm}(x)$ must have simple poles at the branch
points. Moreover, because $\Omega_{\pm}(x)$ obey the boundary
conditions given in (\ref{w1barbc}) - (\ref{wpmconst}) with the
asymptotic behaviors given by (\ref{w1ps1sasy}),
$\Omega_{+}(x)\Omega_{-}(x)$ is even in $x$ and with the simple
poles at $\pm a$ and $\pm b$, we can write it in its most general
form as
\begin{equation}
\Omega_{+}(x)\Omega_{-}(x)=\frac{S^2}{3}\frac{x^2-e^2}
        {(x^2-a^2)(x^2-b^2)}\,,
        \
\label{wpmprop1}
\end{equation}
where $e$ is a constant and $\Omega_{+}(x)\Omega_{-}(x)$ has two
zeros at $x=\pm e$. We will choose $x=+e$ to be a zero of
$\Omega_{+}(x)$ and $x=-e$ to be a zero of $\Omega_{-}(x)$.
Following \cite{EKZ}, it is convenient to define functions that will
simplify our notations,
\begin{equation}
g_{\pm}(x)=\frac{\sqrt{(x^2-a^2)(x^2-b^2)}\pm\frac{x}{e}
\sqrt{(e^2-a^2)(e^2-b^2)}} {x^2-e^2}.\label{gpm1}
\end{equation}
The functions $\Omega_{\pm}(x)$ that satisfy the above properties
 can then be written in their most general forms as
\begin{equation}
\Omega_{\pm}(x)=-\frac{i}{\sqrt{(x^2-a^2)(x^2-b^2)}}
\Bigl((x^2-e^2)(cg_{\mp}(x)\pm d x)\Bigl)^{1/3}, \label{wpgs}
\end{equation}
where $c$ and $d$ are constants. Putting (\ref{wpgs}) in
(\ref{wpmprop1}), we obtain
\begin{eqnarray}
\frac{a^2 b^2}{e^2}c^{2}-c^2 x^2-d^2  e^2
x^2-2\frac{cd}{e}\sqrt{(e^2-a^2)(e^2-b^2)}\,x^2 +d^2
x^4-\frac{1}{27}S^6(x^2-e^2)^2=0. \label{abeeqn}
\end{eqnarray}
The constants $d$, $c$ and $e$ are expressed in terms of $a$ and $b$
using (\ref{abeeqn}) at $x=\infty$, $0$ and $e$ which give
\begin{equation}
d=\frac{1}{3\sqrt{3}} S^{3}\,,\label{cdpol3}
\end{equation}
\begin{equation}
c=-\frac{2}{3\sqrt{3}}S^3
\frac{\sqrt{(e^2-a^2)(e^2-b^2)}}{e-a^2b^2/e^3},\label{cdpol4}
\end{equation}
and
\begin{equation}
e^{4}+2ab\sqrt{(e^2-a^2)(e^2-b^2)}-a^2b^2=0, \label{cdpol2}
\end{equation}
where the appropriate signs are chosen such that the desired
asymptotic behaviors are produced.

It also follows from (\ref{w1spmdec}) and the constraint that
$w_{s\pm}(x)$ and $\Omega_{\pm}(x)$ satisfy the same boundary
conditions that
\begin{equation}
h_{\pm}(x-i0)= h_{\mp}(x+i0)\quad \mathrm{for} \, x\in [a,\,b]\,
\mathrm{and}\,x\in [-b,\,-a]. \label{hpmarasy}
\end{equation}
Thus $h_{+}(x)+h_{-}(x)$ is regular while $h_{+}(x)-h_{-}(x)$ has
square root branch cuts across $x\in [a,\,b]$ and $x\in [-b,\,-a]$.
Because $w_{\pm}(x)$ and $\Omega_{\pm}(x)$ have the asymptotic
behaviors given in (\ref{w1ps1sasy}) and $w_{s\pm}(x)$ have the
asymptotic behavior $\sim x^{n}$, $h_{\pm}(x)$ must have the large
$x$ behavior $\sim x^{n+1}$. We can write $h_{\pm}(x)$ that
satisfies these constraints in the most general form as
\begin{equation}
h_{\pm}(x)=\sqrt{(x^2-a^2)(x^2-b^2)} \Bigl(A(x^2)g_{\pm}(x) \pm
xB(x^2)\Bigr),\label{hgenf}
\end{equation}
where $A(x^2)$ and $B(x^2)$ are even polynomials of degree at most
$n-2$ and $n-4$ respectively if $n$ is even and each of degree at
most $n-3$ if $n$ is odd. We then obtain $w_{s\pm}(x)$ using
(\ref{wpgs}) and (\ref{hgenf}) with (\ref{cdpol3}) - (\ref{cdpol2})
in (\ref{w1spmdec}),
\begin{eqnarray}
w_{s\pm}(x)=-i\frac{1}{\sqrt{3}}S\Bigl((x^2-e^2)(\frac{e^3}{ab}
g_{\mp}(x)\pm x)\Bigl)^{1/3} \Bigl(A(x^2)g_{\pm}(x) \pm
xB(x^2)\Bigr), \label{wispm-2}
\end{eqnarray}
where $A(x^2)$, $B(x^2)$ and the constants $a$, $b$ and $e$ are
calculated putting (\ref{a2-19a}) and (\ref{wispm-2}) in
(\ref{wpmrsa}), making use of the asymptotic behaviors
(\ref{w1ps1sasy}), and using the constraint given by (\ref{cdpol2})
for any given tree level superpotentials $W_1(x)$ and $W_2(x)$
symmetric about the origin and the resolvents having separate cuts.
With the resolvents completely determined in terms of the input
parameters of the theory, we have solved the matrix model in the
planar limit.

The polynomials $p(x)$ and $q(x)$ follow from (\ref{wispm-2}) in
(\ref{pwpwm}) and (\ref{qwqwm}), see Appendix \ref{pqpolyn} for more
details,
\begin{equation}
p(x)=\frac{S^2}{3}\Bigl((x^2-\frac{a^2 b^2}{e^2})A(x^2)^2+
(\frac{e^3}{ab}-\frac{ab}{e}) x^2 A(x^2)B(x^2) -(x^4-e^2 x^2)
B(x^2)^2\Bigr) \label{pxpol-1}
\end{equation}
and
\begin{eqnarray}
q(x)&=&\frac{i}{6\sqrt{3}}S^{3}\Bigl[ \Bigl((3\frac{a^2
b^2}{e}-e^3)x^2+2\frac{a^3 b^3}{e^3}\Bigl)A(x^2)^3 +
3\Bigl(2x^4-(\frac{a^2b^2}{e^2}+e^2)x^2\Bigl)A(x^2)^2 B(x^2)
\nonumber\\&&+  3\Bigl((\frac{e^3}{ab}+\frac{a^2b^2}{e})x^4-2abex^2
\Bigl)A(x^2) B(x^2)^2 + \Bigl(2x^6+(\frac{e^6}{a^2 b^2}-3e^2)x^4
\Bigl)B(x^2)^3 \Bigr]. \label{qxpol-1}
\end{eqnarray}
With $p(x)$ and $q(x)$ in hand, we have found the explicit forms of
the quantum resolution functions putting (\ref{pxpol-1}),
(\ref{qxpol-1}) and (\ref{a2-19a}) in (\ref{quantum-f1}) and
(\ref{quantum-g1}).

The final result for the spectral curve that describes quantum
geometry follows from (\ref{a2-19a}), (\ref{quantum-sc1b}) -
(\ref{quantum-g}), (\ref{pxpol-1}) and (\ref{qxpol-1}),
\begin{eqnarray}
&&(y-\frac{1}{3}(2W'_1(x)+W'_2(x))(y
+\frac{1}{3}(W'_1(x)+2W'_2(x))\nonumber\\
&&(y+\frac{1}{3}(W'_1(x)-W'_2(x))-f(x)\,y-g(x)=0,
\label{quantum-geomf}
\end{eqnarray}
where $f(x)$ and $g(x)$ are given by
\begin{eqnarray}
f(x)&=& S^2 \Bigl((x^2-\frac{a^2 b^2}{e^2})A(x^2)^2+
(\frac{e^3}{ab}-\frac{ab}{e}) x^2 A(x^2)B(x^2) -(x^4-e^2 x^2)
B(x^2)^2\Bigr) \nonumber\\&&-\frac{1}{3}(W_{1}'(x)^2+
W_{2}'(x)^2+W_{1}'(x)W_{2}'(x)),\label{quantum-ff1}
\end{eqnarray}
\begin{eqnarray}
g(x)&=& \frac{i}{3\sqrt{3}}S^{3}\Bigl[ \Bigl((3\frac{a^2
b^2}{e}-e^3)x^2+2\frac{a^3 b^3}{e^3}\Bigl)A(x^2)^3 +
3\Bigl(2x^4-(\frac{a^2b^2}{e^2}+e^2)x^2\Bigl)A(x^2)^2 B(x^2)
\nonumber\\&&+ 3 \Bigl((\frac{e^3}{ab}+\frac{a^2b^2}{e})x^4-2abex^2
\Bigl)A(x^2) B(x^2)^2 + \Bigl(2x^6+(\frac{e^6}{a^2 b^2}-3e^2)x^4
\Bigl)B(x^2)^3
\Bigr]\nonumber\\&&-\frac{1}{27}(2W_{1}'(x)^{3}-2W_{2}'(x)^{3}
+3W_{1}'(x)^{2}W_{2}'(x)-3W_{1}'(x)W_{2}'(x)^{2}).\label{quantum-gf1}
\end{eqnarray}
The even polynomials $A(x^2)$ and $B(x^2)$ and the constants $a$,
$b$ and $e$ in (\ref{quantum-ff1}) and (\ref{quantum-gf1}) are
completely determined for any given symmetric tree level
superpotentials such that the branch cuts of $w_1(x)$ and $w_2(x)$
are disconnected and on opposite sides of the origin making use of
the relation given by (\ref{cdpol2}) and the asymptotic behaviors of
$w_{\pm}(x)$ given by (\ref{w1ps1sasy}).

Let us now apply our results to the simple case of symmetric
quadratic potentials,
\begin{equation}
W_{1}(x)=\frac{1}{2}m x^{2}-\alpha x \quad \mathrm{and} \quad
W_{2}(x)=\frac{1}{2}m x^{2}+\alpha x\label{a2-18}
\end{equation}
where $m$ and $\alpha$ are constants such that $w_{1}(x)$ and
$w_{2}(x)$ have non overlapping branch cuts so that the gauge
symmetry is unbroken in the low energy theory. The regular parts of
the resolvents $w_{1r}(x)$ and $w_{2r}(x)$ follow from (\ref{a2-18})
in (\ref{a2-19a}),
\begin{equation}
w_{1r}(x)= -\frac{1}{3S}\Bigl(3m x-\alpha \Bigr),\quad\quad
w_{2r}(x)= -\frac{1}{3S}\Bigl(3m x+\alpha \Bigr). \label{a2-19aa}
\end{equation}
Now (\ref{a2-19aa}) in (\ref{w1ps1s}) with $w_{is}$ and  $w_{s\pm}$
replaced by $w_{ir}$ and $w_{r\pm}$ give
\begin{equation}
w_{r\pm}(x)= \mp\frac{i}{\sqrt{3}}m x-\frac{1}{3}\alpha.
\label{w1rpab}
\end{equation}
In this case, the asymptotic behaviors of $w_{\pm}(x)$ require that
$B(x^2)=0$ and $A(x^2)=A$, where $A$ is a constant, which with
(\ref{wispm-2}) and (\ref{w1rpab}) in (\ref{wpmrsa}) give
\begin{eqnarray}
w_{\pm}(x)=\mp\frac{i}{\sqrt{3}}m x-\frac{1}{3}\alpha
-i\frac{1}{\sqrt{3}}A S\Bigl((x^2-e^2)(\frac{e^3}{ab}g_{\mp}(x)\pm
x)\Bigl)^{1/3} g_{\pm}(x). \label{wplusall}
\end{eqnarray}
Note that there are a total of four unknown parameters $A$, $a$, $b$
and $e$. Demanding that $w_{\pm}(x)$ in (\ref{wplusall}) obey the
asymptotic large $x$ limits given by (\ref{w1ps1sasy}) gives three
equations and we have one additional constraint among $a$, $b$ and
$e$ given by (\ref{cdpol2}). The asymptotic limits (\ref{w1ps1sasy})
on (\ref{wplusall}) give the following relations
\begin{equation}
A=-\frac{m}{S}, \label{w1pc1}
\end{equation}
\begin{equation}
e=-i\sqrt{3}\frac{ m}{\alpha}a b, \label{w1pc2}
\end{equation}
and
\begin{equation}
m(a^2+b^2)+4\frac{m^3}{\alpha^2}a^2 b^2-6\frac{m^7}{\alpha^6}a^4 b^4
=2S. \label{w1pc3}
\end{equation}
Combining (\ref{cdpol2}) and (\ref{w1pc3}), we obtain a simple
expression for the sum of the squares of the locations of the branch
points,
\begin{equation}
a^2+b^2=18\frac{S}{m}+2\frac{\alpha^2}{m^2}. \label{w1pc4}
\end{equation}
Explicit expressions for the locations of the branch points $\pm a$
and $\pm b$ and the constant parameter $e$ are given in Appendix
\ref{appbp}. The functions $f(x)$ and $g(x)$ that parameterize the
quantum resolution of the classical Calabi-Yau singularities are
also given in Appendix \ref{appqrq}. The constants $a$, $b$ and $e$
are all completely determined in terms of the parameters of the
theory $m$, $\alpha$ and $S$ through (\ref{aquadf}), (\ref{bquadf})
and (\ref{equadf}). Note also that $a$, $b$ and $e$ have nice
relations. The constants $a$ and $b$ are real for real $S$, $m$ and
$\alpha$ as we demanded and the magnitudes of $a$ and $b$ become
larger as the critical points of the potentials $x=\pm \alpha/m$ get
further away from the origin. Moreover, $e$ is pure imaginary and
nonzero for $a\ne 0$. Our solution describes a theory in which the
gauge symmetry in the low energy theory is preserved and $\alpha/m$
is such that the two branch cuts, one from $w_{1}(x)$ and the other
from $w_{2}(x)$ are disconnected with $b>a>0$. As the parameters of
the theory $\alpha$, $m$ and $S$ are varied, the locations of the
branch points move on the real axis of $x$ and this is related to a
movement of $e$ on the imaginary axis of $x$.

\section{Conclusion}\label{concl}

In conclusion, matrix models in combination with supersymmetric
gauge theories provide very powerful tools that allow us to study
exact nonperturbative physics for systems involving quite general
tree level superpotentials where symmetries and holomorphy alone are
not enough. On the other hand, a class of supersymmetric gauge
theories with tree level superpotentials can be geometrically
engineered in type $IIA$ and type $IIB$ string theories by wrapping
D-branes over various cycles of Calabi-Yau threefolds. The
singularities in the classical Calabi-Yau geometry are resolved by
quantum effects. In this note we have used the combined power of
supersymmetry and matrix models to construct the exact quantum
Calabi-Yau  geometries associated to $\mathcal{N}=1$ supersymmetric
$A_{2}$ type $U(N)\times U(N)$ gauge theories with quite general
symmetric tree level polynomial superpotentials of any degree. Even
though our interest in this note was the construction of the quantum
geometries, our exact results could be used to compute the free
energies in the planar limit and the exact nonperturbative dynamical
superpotentials of $A_2$.

\section*{Acknowledgements}

Part of this research was supported by the Department of Energy
under grant number DE-FG02-91ER40654.

\appendix

\setcounter{equation}{0}
\section{Multi-matrix free energy integral}\label{appfii}

Here we give the derivation of (\ref{f0rs}) which gives the
multi-matrix planar free energy in terms of simple single integrals.
First the free energy in the large $N$ limit can be read off from
the effective action given by (\ref{eq:zpq-3}) and it is
\begin{eqnarray}
\mathcal{F}_{0} & = & \sum_{i}S_{i}\int d\lambda\rho_{i}
(\lambda)W_{i}(\lambda)-\sum_{i}S_{i}^{2}\int\int d\lambda
d\mu\rho_{i}(\lambda)\rho_{i}(\mu)\mathrm{log}|\lambda-\mu|\nonumber \\
 &  & +\sum_{i<j}|s_{ij}|S_{i}S_{j}\int\int d\lambda d\mu\rho_{i}
 (\lambda)\rho_{j}(\mu)\mathrm{log}|\lambda-\mu|.\label{f0r}
 \end{eqnarray}
Our notation is such that in the planar limit the free energy is
related to the partition function via
$Z=e^{-\mathcal{F}_{0}/g_{s}^2}$. See \cite{H1} for more about our
notation. The first two terms come from each cut separately while
the last term is due to interactions between different sets. Taking
the large $N$ limit of (\ref{eq:zpq-4}) and integrating the equation
of motion over $\lambda$  gives
\begin{eqnarray}
S_{i}\int
d\mu\rho_{i}(\mu)\log|\lambda-\mu|&=&\frac{1}{2}W_{i}(\lambda)
+S_{i}\int d\mu
\rho_{i}(\mu)\log{|\mu|}+\frac{1}{2}\sum_{j}|s_{ij}|S_{j}\int
d\mu\rho_{j}(\mu)\log|\lambda-\mu|\nonumber \\
&&-\frac{1}{2}\sum_{j}|s_{ij}|S_{j} \int
d\mu\rho_{j}(\mu)\log|\mu|.\label{appfin1}
\end{eqnarray}
Substituting (\ref{appfin1}) for $S_{i}\int
d\mu\rho_{i}(\mu)\log|\lambda-\mu|$ in the second term of
(\ref{f0r}), remembering that the eigenvalue densities are
normalized such that $\int \rho_{i}(\lambda)d\lambda=1$, and
simplifying we obtain
\begin{equation} \mathcal{F}_{0}=
\frac{1}{2}\sum_{i}S_{i}\int d\lambda\rho_{i}
(\lambda)W_{i}(\lambda)-\frac{1}{2}\sum_{i,j}C_{ij}S_{i}S_{j}\int
d\lambda\rho_{i} (\lambda)\mathrm{log}|\lambda|\,\label{appf0rs}
 \end{equation}
where $C_{ij}=2\delta_{ij}-|s_{ij}|$ is the Cartan matrix.

\setcounter{equation}{0}
\section{The quadratic and the cubic equations}\label{appqce}

Here we give a derivation of the quadratic and the cubic equations.
The equations of motion of the eigenvalues for the $U(N_1)\times
U(N_2)$ matrix model are
\begin{equation}
\frac{1}{g_{s}}W'_{1}(\lambda_{1,I})-2\sum_{J\ne
I}\frac{1}{\lambda_{1,I}-\lambda_{1,J}}+\sum_{J}\frac{1}
{\lambda_{1,I}-\lambda_{2,J}}=0,\label{aq1}\end{equation}
\begin{equation}
\frac{1}{g_{s}}W'_{2}(\lambda_{2,I})-2\sum_{J\ne
I}\frac{1}{\lambda_{2,I}-\lambda_{2,J}}+\sum_{J}\frac{1}
{\lambda_{2,I}-\lambda_{1,J}}=0.\label{aq2}\end{equation} Squaring
the resolvent $w_{1}(x)$ defined by (\ref{eq:zpq-5}),
\begin{equation}
w_{1}(x)^{2}=\frac{1}{N_{1}}w_{1}'(x)+\frac{2}{N_{1}^{2}}
\sum_{I}\frac{1}{\lambda_{1,I}-x}\sum_{J\ne
I}\frac{1}{\lambda_{1,I}-\lambda_{1,J}}.\label{aq3}\end{equation}
Using the first equation of motion (\ref{aq1}) to substitute for
$\sum_{J\ne I}\frac{1}{\lambda_{1,I}-\lambda_{1,J}}$ in (\ref{aq3}),
\begin{equation}
w_{1}(x)^{2}=\frac{1}{N_{1}}w_{1}'(x)-\frac{1}{N_{1}S_{1}}
\sum_{I}\frac{W_{1}'(\lambda_{1I})}{\lambda_{1,I}-x}-\frac{1}
{N_{1}^{2}}\sum_{I,J}\frac{1}{\lambda_{1,I}-x}\frac{1}
{\lambda_{1,I}-\lambda_{2,J}}.\label{aq4}
\end{equation}
But
\begin{eqnarray}
\frac{1}{N_{1}^{2}}\sum_{I,J}\frac{1}{\lambda_{1,I}-x}\frac{1}
{\lambda_{1,I}-\lambda_{2,J}} =
-\frac{N_{2}}{N1}w_{1}(x)w_{2}(x)+\frac{1}{N_{1}^{2}}\sum_{J}\frac{1}
 {\lambda_{2,J}-x}\sum_{I}\frac{1}{\lambda_{1,I}-\lambda_{2,J}}.\label{aq5}
\end{eqnarray}
Using (\ref{aq2}) to substitute for $\sum_{I\ne
J}\frac{1}{\lambda_{1,I}-\lambda_{2,J}}$ in (\ref{aq5}),
\begin{eqnarray}
\frac{1}{N_{1}^{2}}\sum_{I,J}\frac{1}{\lambda_{1,I}-x}\frac{1}
{\lambda_{1,I}-\lambda_{2,J}} & = & -\frac{N_{2}}{N1}w_{1}
(x)w_{2}(x)+\frac{1}{N_{1}S_{1}}\sum_{J}\frac{W_{2}'(\lambda_{2J})}
{\lambda_{2,J}-x}\nonumber \\
 &  & -\frac{2}{N_{1}^{2}}\sum_{J}\frac{1}{\lambda_{2,J}-x}
 \sum_{I\ne J}\frac{1}{\lambda_{2,I}-\lambda_{2,I}}.\label{aq6}
 \end{eqnarray}
Next squaring $w_{2}(x)$ we write the last term in (\ref{aq5}) as
\begin{equation}
\frac{2}{N_{1}^{2}}\sum_{J}\frac{1}{\lambda_{2,J}-x}\sum_{I\ne
J}\frac{1}{\lambda_{2,I}-\lambda_{2,I}}=\frac{N_{2}^{2}}{N_{1}^{2}}
w_{2}(x)^{2}-\frac{N_{2}}{N_{1}^{2}}w_{2}'(x).\label{aq7}
\end{equation}
Using (\ref{aq6}) and (\ref{aq7}) in (\ref{aq4}),
\begin{eqnarray}
w_{1}(x)^{2}-\frac{1}{N_{1}}w_{1}'(x)+\frac{1}{N_{1}S_{1}}
\sum_{I}\frac{W_{1}'(\lambda_{1I})}{\lambda_{1,I}-x}
-\frac{N_{2}}{N1}w_{1}(x)w_{2}(x)\nonumber \\
+\frac{1}{N_{1}S_{1}}\sum_{J}\frac{W_{2}'(\lambda_{2J})}
{\lambda_{2,J}-x}+\frac{N_{2}^{2}}{N_{1}^{2}}w_{2}(x)^{2}
-\frac{N_{2}}{N_{1}^{2}}w_{2}'(x)=0\label{aq8}
\end{eqnarray}
But we can write
\begin{equation}
\frac{1}{N_{1}S_{1}}\sum_{I}\frac{W_{1}'(\lambda_{1I})}
{\lambda_{1,I}-x}=\frac{1}{S_{1}}W_{1}'(x)w_{1}(x)+\frac{1}
{S_{1}^{2}}f_{1}(x)\label{aq9}
\end{equation}
and
\begin{equation}
\frac{1}{N_{1}S_{1}}\sum_{I}\frac{W_{2}'(\lambda_{2I})}
{\lambda_{2,I}-x}=\frac{N_{2}}{N_{1}S_{1}}W_{2}'(x)w_{2}(x)
+\frac{N_{2}}{S_{1}S_{2}N_{1}}f_{2}(x),\label{aq10}
\end{equation}
where
\begin{equation}
f_{i}(x)\equiv\frac{S_{i}}{N_{i}}\sum_{I}\frac{W_{i}'(x)
-W_{i}'(\lambda_{iI})}{x-\lambda_{i,I}}.\label{apaq11}
\end{equation}
Putting (\ref{aq9}) and (\ref{aq10}) in (\ref{aq8}) and discarding
the $\frac{1}{N_{i}}w_{i}'(x)$ terms in the large $N$ limit gives
the quadratic equation
\begin{eqnarray}
S_{1}^{2}w_{1}(x)^{2}+S_{2}^{2}w_{2}(x)^{2}
-S_{1}S_{2}w_{1}(x)w_{2}(x)+S_{1}W_{1}'(x)w_{1}(x)\nonumber \\
 +S_{2}W_{2}'(x)w_{2}(x)
+f_{1}(x) +f_{2}(x)=0.\label{apw12quad}
\end{eqnarray}
Repeating a similar procedure as above starting with $w_{1}(x)^2
w_{2}(x)$ one obtains the cubic equation
\begin{eqnarray}
S_{1}^2 S_{2} w_{1}(x)^{2}w_{2}(x)-S_{1} S_{2}^{2} w_{1}(x)
w_{2}(x)^{2}+S_{1}^{2} w_{1}(x)^2W_{1}'(x)+S_{1}w_{1}(x)W_{1}'(x)^2
+f_{1}(x)W_{1}'(x)\nonumber\\
-g_{1}(x)-S_{2}^{2}w_{2}(x)^2
W_{2}'(x)-S_{2}w_{2}(x)W_{2}'(x)^2-f_{2}(x)W_{2}'(x)+g_{2}(x)=0,
\label{apw12cub}
\end{eqnarray}
where
\begin{equation}
g_{1}(x)\equiv\frac{S_{1}S_{2}}{N_{1}N_{2}}\sum_{I,J}\frac{W_{1}'(x)
-W_{1}'(\lambda_{1I})}{(\lambda_{1,I}-\lambda_{2,J})(x-\lambda_{1,I})},\quad
g_{2}(x)\equiv\frac{S_{1}S_{2}}{N_{1}N_{2}}\sum_{I,J}\frac{W_{2}'(x)
-W_{2}'(\lambda_{2I})}{(\lambda_{2,I}-\lambda_{1,J})(x-\lambda_{2,I})}.
\label{apaq11b}
\end{equation}

\setcounter{equation}{0}
\section{The polynomials $p(x)$ and $q(x)$}\label{pqpolyn}

Here we prove that the $p(x)$ and $q(x)$ that follow from the
singular parts of the quantum resolvents given by (\ref{wispm-2})
are indeed polynomials and their expressions are given. First the
polynomial $p(x)$ is easily constructed. Using the decomposition
given in (\ref{w1spmdec}) with (\ref{wpmprop1}) and (\ref{hgenf}) in
(\ref{pwpwm}) gives
\begin{equation}
p(x)=\frac{S^2}{3}(x^2-e^2)\Bigl(g_{+}(x)g_{-}(x)A(x^2)^2
-x(g_{+}(x)-g_{-}(x))A(x^2)B(x^2)-x^2B(x^2)^2\Bigr)\label{ppol-ap1}
\end{equation}
But (\ref{gpm1}) gives
\begin{equation}
(x^2-e^2)(g_{+}(x)g_{-}(x))=x^2-\frac{a^2 b^2}{e^2}\quad
\mathrm{and} \quad (x^2-e^2)(g_{+}(x)-g_{-}(x))=
2\frac{x}{e}\sqrt{(e^2-a^2)(e^2-b^2)}.\label{ppol-ap2}
\end{equation}
Moreover, (\ref{cdpol2}) gives
\begin{equation}
\sqrt{(e^2-a^2)(e^2-b^2)}=\frac{1}{2}(ab-\frac{e^4}{ab}).
\label{ppol-ap3}
\end{equation}
Putting (\ref{ppol-ap2}) and (\ref{ppol-ap3}) in (\ref{ppol-ap1}),
\begin{equation}
p(x)=\frac{S^2}{3}\Bigl((x^2-\frac{a^2 b^2}{e^2})A(x^2)^2+
(\frac{e^3}{ab}-\frac{ab}{e})x^2 A(x^2)B(x^2) -(x^4-e^2 x^2)
B(x^2)^2\Bigr). \label{pxpol-2}
\end{equation}

For $q(x)$ we start with (\ref{wispm-2}) in (\ref{qwqwm}) which
gives
\begin{eqnarray}
q(x)&=&\frac{i}{6\sqrt{3}}S^{3}(x^2-e^2)\Bigl[
\Bigl(\frac{e^3}{ab}g_{+}(x)g_{-}(x)(g_{+}(x)^2+g_{-}(x)^2) +
x(g_{+}(x)^3-g_{-}(x)^3)\Bigl)A(x^2)^3 \nonumber\\&&+ 3x
\Bigl(\frac{e^3}{ab}g_{+}(x)g_{-}(x)(g_{+}(x)-g_{-}(x)) +x
(g_{+}(x)^2+g_{-}(x)^2) \Bigl)A(x^2)^2 B(x^2) \nonumber\\&&+ 3x^2
\Bigl(2\frac{e^3}{ab}g_{+}(x)g_{-}(x) + x (g_{+}(x)-g_{-}(x))
\Bigl)A(x^2) B(x^2)^2 \nonumber\\&&- x^3
\Bigl(\frac{e^3}{ab}(g_{+}(x)-g_{-}(x)) -2 x \Bigl)B(x^2)^3 \Bigr].
\label{apqxpol-1}
\end{eqnarray}
After some algebra involving $g_{\pm}(x)$ and also making use of the
constraint given by (\ref{cdpol2}), we obtain
\begin{eqnarray}
q(x)&=&\frac{i}{6\sqrt{3}}S^{3}\Bigl[ \Bigl((3\frac{a^2
b^2}{e}-e^3)x^2+2\frac{a^3 b^3}{e^3}\Bigl)A(x^2)^3 +
3\Bigl(2x^4-(\frac{a^2b^2}{e^2}+e^2)x^2\Bigl)A(x^2)^2 B(x^2)
\nonumber\\&&+  3\Bigl((\frac{e^3}{ab}+\frac{a^2b^2}{e})x^4-2abex^2
\Bigl)A(x^2) B(x^2)^2 + \Bigl(2x^6+(\frac{e^6}{a^2 b^2}-3e^2)x^4
\Bigl)B(x^2)^3 \Bigr]. \label{apqxpol-2}
\end{eqnarray}
Thus the quantum resolvents indeed produce polynomials $p(x)$ and
$q(x)$ in terms of two other even polynomials $A(x^2)$ and $B(x^2)$
which are specific to the tree level superpotentials $W_{1}(x)$ and
$W_{2}(x)$.

\setcounter{equation}{0}
\section{Locations of branch points}\label{appbp}
Here we give explicit expressions for the locations of the branch
points and the imaginary constant parameter $e$ for the example of
symmetric quadratic tree level superpotentials we discussed at the
end of Section \ref{sqga2}. First (\ref{cdpol2}) gives
\begin{equation}
e^8-6a^2b^2e^4+4a^2b^2(a^2+b^2)e^2-3a^4b^4=0.\label{ebsq1}
\end{equation}
Putting (\ref{w1pc2}) in (\ref{ebsq1}),
\begin{equation}
27\frac{m^8}{\alpha^8}a^4 b^4- 18\frac{m^4}{\alpha^4} a^2 b^2
-4\frac{m^2}{\alpha^2}(a^2+b^2)- 1 =0.\label{ebsq2}
\end{equation}
Finally, solving (\ref{w1pc4}) and (\ref{ebsq2}) simultaneously and
remembering that the phases and magnitudes of $\pm a$ and $\pm b$
are such that $b>a>0$, we obtain
\begin{equation}
a=\sqrt{9\frac{S}{m}+\frac{\alpha^2}{m^2}
-\sqrt{\frac{2}{3}\frac{\alpha^4}{m^4} +18\frac{\alpha^2 S}{m^3} +
81\frac{S^2}{m^2}-\sqrt{\frac{4}{9}\frac{\alpha^8}{m^8}
+\frac{8}{3}\frac{\alpha^6 S}{m^7}}}}\label{aquadf}
\end{equation}
and
\begin{equation}
b=\sqrt{9\frac{S}{m}+\frac{\alpha^2}{m^2}
+\sqrt{\frac{2}{3}\frac{\alpha^4}{m^4} +18\frac{\alpha^2 S}{m^3} +
81\frac{S^2}{m^2}-\sqrt{\frac{4}{9}\frac{\alpha^8}{m^8}
+\frac{8}{3}\frac{\alpha^6 S}{m^7}}}}.\label{bquadf}
\end{equation}
Putting (\ref{aquadf}) and (\ref{bquadf}) in (\ref{w1pc2}), we also
obtain $e$,
\begin{equation}
e=-i \sqrt{\frac{\alpha^2}{m^2} +\sqrt{4 \frac{\alpha^4}{m^4}
+24\frac{\alpha^2 S}{m^3}}}\label{equadf}
\end{equation}
 Note that $a$, $b$ and $e$ are all completely
determined in terms of the parameters of the theory $m$, $\alpha$
and $S$ through (\ref{aquadf}), (\ref{bquadf}) and (\ref{equadf}).
The locations of the branch points $\pm a$ and $\pm b$, the branch
cuts, and the pure imaginary constant $e$ are schematically shown in
Figure \ref{fig-abe}. As the parameters $\alpha/m$ and $S/m$ are
increased, the branch cuts and the imaginary parameter $e$ get
further away from the origin.
{\figsize\begin{figure}[htb]
{\centerline{ \epsfbox{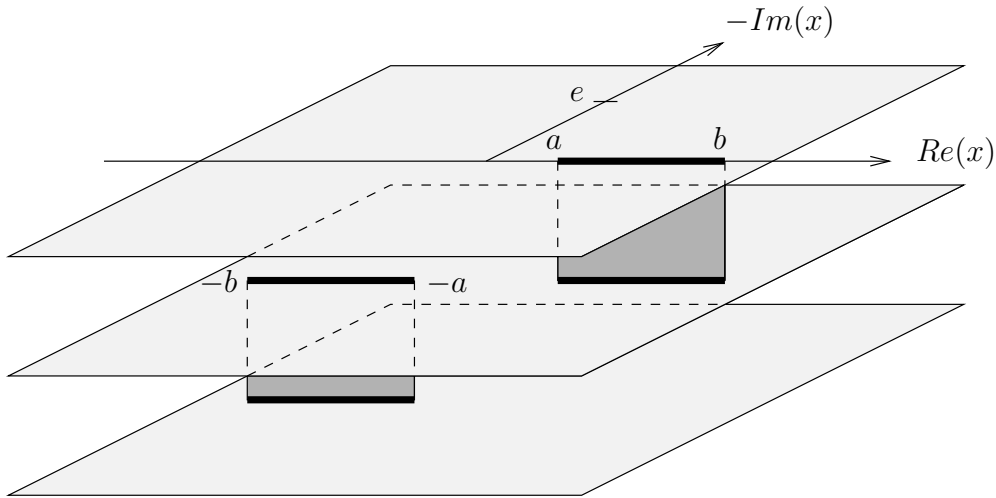}}}
\caption{\figsize\sf\label{fig-abe}The three sheets, the two branch
cuts, and $e$. The branch points are $\pm a$ and $\pm b$. The cuts
are on the real axis of $x$ and the one across $[a,\,b]$ is for
$w_{1}(x)$ and it joins the first and the second sheets, the cut
across $[-b,\,-a]$ is for $w_{2}(x)$ and it joins the second and the
third sheets. The constant $e$ is pure imaginary and its relation to
$a$ and $b$ is given by (\ref{w1pc2}). }\end{figure}}

\setcounter{equation}{0}
\section{Quantum resolution functions}\label{appqrq}
Here we write down the functions that describe the quantum
resolution of the classical singularities of the Calabi-Yau geometry
for the simple example of symmetric quadratic tree level
superpotentials we applied our results at the end of Section
\ref{sqga2}. First $p(x)$ and $q(x)$ are obtained using $A$, $e$,
$a$ and $b$ found in (\ref{w1pc1}), (\ref{w1pc2}), (\ref{aquadf})
and (\ref{bquadf}) in (\ref{pxpol-1}) and (\ref{qxpol-1}),
\begin{equation}
p(x)=\frac{1}{3}m^2 x^2+\frac{1}{9}\alpha^2,\label{pquad2}
\end{equation}
\begin{equation}
q(x)=-\frac{1}{3\sqrt{3}}\sqrt{\alpha^3(\alpha^2+6mS)
(\alpha+2\sqrt{\alpha^2+6mS})}\,x^2-\frac{1}{27}\alpha^3.
\label{qquad2}
\end{equation}
Putting (\ref{a2-18}), (\ref{pquad2}) and (\ref{qquad2}) in
(\ref{quantum-f}) and (\ref{quantum-g}), we obtain
\begin{equation}
f(x)=0,\label{ffquad2}
\end{equation}
\begin{equation}
g(x)=\Bigl(\frac{2}{3}\alpha m^2
-\frac{2}{3\sqrt{3}}\sqrt{\alpha^3(\alpha^2+6mS)
(\alpha+2\sqrt{\alpha^2+6mS})}\,\Bigr)x^2-\frac{4}{27}\alpha^3.
\label{gquad2}
\end{equation}

\end{document}